# Analyzing melts and fluids from ab initio molecular dynamics simulations with the UMD package

**AUTHORS:**

Razvan Caracas[1,2], Anais Kobsch[1], Natalia V. Solomatova[1], Zhi Li[1], Francois Soubiran[1,3], Jean-Alexis Hernandez [1,2]

[1] CNRS, Ecole Normale Supérieure de Lyon, Laboratory of Geology of Lyon UMR5276, Lyon, France

[2] University of Oslo, Centre for Earth Evolution and Dynamics (CEED), Oslo, Norway

[3] CEA, DAM, DIF, 91297 Arpajon, France

Email addresses of co-authors:

Razvan Caracas (razvan.caracas@ens-lyon.fr; razvan.caracas@geo.uio.no, razvan.caracas@gmail.com)

Anais Kobsch (astranais@gmail.com)

Natalia V. Solomatova (nsolomat@gmail.com)

Zhi Li (zhi.li@ens-lyon.org)

Francois Soubiran (francois.soubiran@ens-lyon.org)

Jean-Alexis Hernandez (jeanalexis.hernandez@gmail.com)

Corresponding author:

Razvan Caracas (razvan.caracas@gmail.com)

**ABSTRACT:**

We develop a Python-based open-source package to analyze the results stemming from ab initio molecular-dynamics simulations of fluids. The package is best suited for applications on natural systems, like silicate and oxide melts, water-based fluids, various supercritical fluids. The package is a collection of Python scripts that include two major libraries dealing with file formats and with crystallography. All the scripts are run at the command line. We propose a simplified format to store the atomic trajectories and relevant thermodynamic information of the simulations, which is saved in UMD files, standing for Universal Molecular Dynamics. The UMD package allows the computation of a series of structural, transport and thermodynamic properties. Starting with the pair-distribution function it defines bond lengths, builds an interatomic connectivity matrix, and eventually determines the chemical speciation. Determining the lifetime of the chemical species allows running a full statistical analysis. Then dedicated scripts compute the mean-square displacements for the atoms as well as for the chemical species. The implemented self-correlation analysis of the atomic velocities yields the diffusion coefficients and the vibrational spectrum. The same analysis applied on the stresses yields the viscosity. The package is available via the GitHub website and via its own dedicated page of the ERC IMPACT project as open-access package.

**INTRODUCTION:**

Fluids and melts are active chemical and physical transport vectors in natural environments. The elevated rates of atomic diffusion favor chemical exchanges and reactions, the low viscosity coupled with varying buoyancy favor large mass transfer, and crystal-melt density relations favor layering inside planetary bodies. The absence of a periodic lattice, typical high-temperatures required to reach the molten state, and the difficulty for quenching makes the determination of a series of obvious properties, like density, diffusion, viscosity, extremely challenging. These difficulties make alternative computational methods strong and useful tools for investigating this class of materials.

With the advent of computing power and the availability of supercomputers, two major numerical atomistic simulations techniques are currently employed to study the dynamical state of a non-crystalline atomistic system, Monte Carlo[1] and molecular dynamics (MD)[1,2]. In Monte Carlo simulations the configurational space is randomly sampled; Monte Carlo methods show linear scaling in parallelization if all sampling observations are independent of each other. The quality of the results depends on the quality of the random number generator and the representativeness of the sampling. Monte Carlo methods show linear scaling in parallelization if the sampling is independent of each other. In molecular dynamics (MD) the configurational space is sampled by time-dependent atomic trajectories. Starting from a given configuration, the atomic trajectories are computed by integrating the Newtonian equations of motion. The interatomic forces can be computed using model interatomic potentials (in classical MD) or using first-principles methods (in ab initio, or first-principles, MD). The quality of the results depends on the length of the trajectory and its capability not to be attracted into local minima.

Molecular dynamics simulations contain a plethora of information, all related to the dynamical behavior of the system. Thermodynamic average properties, like internal energy, temperature, and pressure, are rather standard to compute. They can be extracted from the output file(s) of the simulations and averaged, whereas quantities related directly to the movement of the atoms as well as to their mutual relation need to be computed after extraction of the atomic positions and velocities.

Consequently, a lot of effort was dedicated to visualize the results, and various packages became available today, on different platforms, open source or not [Ovito[3], VMD[4], Vesta[5], Travis[6], etc.]. All these visualization tools deal efficiently with interatomic distances, and as such they allow the efficient computation of pair distribution functions and diffusion coefficients. Various groups performing large-scale molecular dynamics simulations have proprietary software to analyze various other properties resulting from the simulations, sometimes in shareware or other forms of limited access to the community, sometimes limited in scope and use to some specific packages. Sophisticated algorithms to extract information about interatomic bonding, geometrical patterns, and thermodynamics are developed and implemented in some of these packages [3-7], etc.

Here we propose **the UMD package** – an open-source package written in Python to analyze the output of molecular dynamics simulations. The UMD package allows for the computation of a wide range of structural, dynamical and thermodynamical properties (Figure 1). The package is available via the GitHub website (https://github.com/rcaracas/UMD_package) and via its own dedicated page (http://moonimpact.eu/umd-package/) of the ERC IMPACT project as open-access package.

To make it universal and easier to handle, our approach is to first extract all the information related to the thermodynamic state and to the atomic trajectories from the output file of the actual molecular-dynamics run. This information is stored in a dedicated file, whose format is independent on the original MD package where the simulation was run. We name these files *umd* files, which stands for Universal Molecular Dynamics. In this way our UMD package can be easily used by any ab initio group with any software, all with a minimum effort of adaptation. The only requirement to use the present package is to write the appropriate parser from the output of the particular MD software into the umd file format, if this is not yet existent. For the time being we provide such parsers for the VASP[8] and the QBox[9] packages.

The umd files are ascii files; typical extension is *umd.dat* but not mandatory. All the analysis components can read ascii file of the umd format, regardless of the actual name extension. However some of the automatic scripts designed to perform fast large-scale statistics over several simulations specifically look for files with the umd.dat extension. Each physical property is expressed on one line. Every line starts with a keyword. In this way the format is highly adaptable and allows for new properties to be added to the umd file, all the while preserving its readability throughout versions. The first 30 lines of the umd file of the simulation of pyrolite at 4.6GPa and 3000K, which is used below in the discussion part, are shown in Figure 2.

[Place Figure 2 here]

All *umd* files contain a header describing the content of the simulation cell: the number of atoms, electrons, and atomic types, as well as details for each atom, such as its type, chemical symbol, number of valence electrons, and its mass. One empty line marks the end of the header, and separates it from the main part of the umd file.

Then each step of the simulation is detailed. First the instantaneous thermodynamic parameters are given, each on a different line, specifying (i) the name of the parameter, like energy, stresses, equivalent hydrostatic pressure, density, volume, lattice parameters, etc, (ii) its value(s), and (iii) its units. A table describing the atoms comes next. A header line gives the different measures, like Cartesian positions, velocities, charges, etc., and their units. Then each atom is detailed on one line. By group of three, corresponding to the three *x, y, z* axes, there are: the reduced positions, the Cartesian positions folded into the simulation cell, the Cartesian positions (that properly take into account the fact that atoms can traverse several unit cells during a simulation), the atomic velocities, and the atomic forces. The last two entries are scalars: charge and magnetic moment.

There are two major libraries that ensure the proper functioning of the entire package. The *umd_process.py* library deals with the umd files, like reading and printing. The *crystallography.py* library deals with all the information related to the actual atomic structure. The underlying philosophy of the *crystallography.py* library is to treat the lattice as a vectorial space. The unit cell parameters together with their orientation represent the basis vectors. The space has a series of scalar attributes (specific volume, density, temperature, and specific number of atoms), thermodynamic properties (internal energy, pressure, heat capacity, etc.), and a series of tensorial properties (stress and elasticity). Atoms populate this space. The Lattice class defines this ensemble, alongside various few short calculations, like specific volume, density, obtaining the reciprocal lattice from the direct one, etc. The Atoms class defines the atoms. They are characterized by a series of scalar properties (name, symbol, mass, number of electrons, etc.) and a series of vectorial properties (the position in space, either relative to the vectorial basis described in the Lattice class, or relative to universal Cartesian coordinates, velocities, forces, etc.). Apart from these two classes, the *crystallography.py* library contains a series of functions to perform a variety of tests and calculations, such as atomic distances, or cell multiplication. The periodic table of the elements is also included as a dictionary.

The various components of the umd package write several output files. As a general rule, they are all ascii files, all their entries separated by tabs, and they are made as self-explanatory as possible, for example always clearly indicating what physical property is written and what are its units. The *umd.dat* files fully comply with this rule.

**PROTOCOL:**

**1. The analysis of the molecular-dynamics runs is done in several steps.**

**1.1. Each step has one or more dedicated Python scripts that address that specific set of physical properties. All the scripts are run at the command line, and they all employ a series of flags, which are as consistent as possible from one script to another. The flags, their meaning and default values are all summarized in Table 1.**

1.2. **Start with transforming the output of the MD simulation into a UMD file**.

1.2.1. If the MD simulations was done in VASP, then at the command line type:

**VaspParser.py -f <OUTCAR_filename> -i <InitialStep>**

where –f flag defines the name of the VASP OUTCAR file, and the –i the thermalization length.

NOTE: the initial step, defined by –i allows for discarding the first steps of the simulations, which represent the thermalization. In a typical molecular-dynamics run the first part to the calculation represents the thermalization – the time it takes the system for all atoms to describe a Gaussian-like distribution of the temperature, and for the entire system to exhibit fluctuations of the temperature, pressure, energy, etc. around equilibrium values. This thermalization part of the simulation should not be taken into account when analyzing statistical properties of the fluid.

1.3. The UMD files can be transformed into various other formats. For example to facilitate visualization on various other packages, the *.umd* files can be transformed into *.xyz* files. At the command line type:

**umd2xyz.py -f <umdfile> -i <InitialStep> -s <Sampling_Frequency>**

where –f defines the name of the *.umd* file, –i defines the thermalization period to be discarded, and –s the frequency of the sampling of the trajectory stored in the *.umd* file. Default values are –i 0 –s 1, meaning no considering all the steps of the simulation, without any being discarded.

1.4. Reverse the *umd* file into VASP-type POSCAR files using the umd2poscar.py script; snapshots of the simulations can be selected with a predefined frequency. At the command line type:

**umd2poscar.py -f <umdfile> -i <InitialStep> -l <LastStep> -s <Sampling_Frequency>**

where –l represents the last step to be transformed into POSCAR file. Default values are -i 0 -l 10000000 -s 1. This value of –l is big enough to cover a typical entire trajectory.

**2. Perform the structural analysis.**

2.1. Compute the radius of the first coordination sphere, as the first minimum of the radial pair distribution function (PDF), $g_{AB}(r)$ (Figure 3). This is the basis for the entire structural analysis of the fluid; the PDF yields the average bonding status of the atoms in the fluid.

NOTE: the radial pair distribution function (PDF), $g_{AB}(r)$ is the average number of atoms of type B at a distance d_AB within a spherical shell of radius r and thickness dr centered on the atoms of type A (Figure 2):

$$g_{AB}(r) = \frac{N_A - 1}{4\pi\rho^2} <\delta(r - r_{AB})>_B$$

with $\rho$ the atomic density, $N_A$ and $N_B$ the number of atoms of type A and B, and $\delta(r - r_{AB})$ the delta function which is equal to 1 if the atoms A and B lie at a distance between $r$ and $r + dr$
The abscissa of the first maximum of $g_{AB}(r)$ gives the highest probability bond length between the atoms of type A and B, which is the closest to an average bond distance that w can determine. The first minimum delimits the extent of the first coordination sphere. Hence the integral over the PDF up to the first minimum gives the average coordination number. The sum of the Fourier transforms of the $g_{AB}(r)$ for all the pairs of atomic types A and B yields the diffraction pattern of the fluid, as obtained experimentally with a diffractometer. However, in reality as oftentimes the high order coordination spheres are missing from the $g_{AB}(r)$, the diffraction pattern cannot be obtained in its entirety.

[Place Figure 3 here]

2.2. Run the gofrs_umd.py script to compute the pair distribution function $g_{AB}(r)$ for all the pairs of atomic types A and B. The output is written in one ascii file, tab-separated, with the extension *gofrs.dat*. At the command line type:

**gofrs_umd.py -f <UMD_filename> -s < Sampling_Frequency > -d <DiscretizationInterval> -i <InitialStep>**

NOTE: the defaults are Sampling_Frequency (the frequency for sampling the trajectory) = 1 step; DiscretizationInterval (for plotting the g(r)) = 0.01 Angstroms; InitialStep (number of steps in the beginning of the trajectory that are discarded) = 0.

2.3. Extract the interatomic bond distances as the radii of the first coordination spheres. For this, identify the position of the first minimum of the $g_{AB}(r)$ functions: plot the gofrs.dat file in a spreadsheet software and search for the maxima and minima for each pair of atoms. Extract the distances of the first minima, i.e. the abscissa, and write them in a separate file, called for example *bonds.input*. Alternatively run one of the provided *analyze_gofr* scripts to identify the maxima and the minima of the $g_{AB}(r)$ functions. Open and look at the automatically generated file called *bonds.input that contains the interatomic bond distances*.

3. **Perform the speciation analysis**.

3.1. This reveals the topology of bonding between the atoms, using the concept of connectivity within graph theory: the atoms are the nodes and the interatomic bonds are the paths. The code needs the interatomic bond distances defined in the *bonds.input* file.

The connectivity matrix is constructed at each time step: two atoms that lie at a distance smaller than the radius of their corresponding first coordination sphere are considered to be bonded, *i.e.* connected. Various atomic networks are built by treating the atoms as nodes in a graph whose connections are defined by this geometric criterion. These networks are the atomic species, and their ensemble defines the atomic speciation in that particular fluid (Figure 4).

3.2. Run the speciation script to obtain the connectivity matrix and obtain the coordination polyhedra or the polymerization. At the command line type:

**speciation_umd.py -f <UMD_filename> -s <Sampling_Frequency> -i <InputFile> -l <MaxLength> -c <Cations> -a <Anions> -m <MinLife> -r <Rings>** where the -i flag gives the file with the interatomic bond distances, which was produced for example in the previous step. Alternatively run the script with one single length for all bonds defined by the -l flag. The -c flag specifies the central atoms, which and the -a flag the ligands. Both central atoms and ligands can be of different types; in this case they must be separated by comma. The -m flag gives the minimum time a species must live to be considered in the analysis. By default this minimum time is zero, all occurrences being counted in the final analysis.

3.2.1. Obtain the coordination polyhedra using the speciation_umd.py script with the flag –r 0, which samples the connectivity graph at the first level. For example, a central atom, denoted as a *cation* may be surrounded by one or more *anions* (Figure 4). The speciation script identifies every single one of the coordination polyhedra. The weighted average of all the coordination polyhedra gives the coordination number, identical to the one obtained from the integration of the PDF.

NOTE: Average coordination numbers in fluids are fractional numbers. This fractionality comes from the average characteristic of the coordination. The definition based on speciation yields a more intuitive and informative representation of the structure of the fluid, where the relative proportions of the different species, *i.e.* coordinations, are quantified.

3.2.2. Obtain the polymerization using the speciation_umd.py script with the flag –r 1, which samples the connectivity graph at all depth levels. The network through the atomic graph has a certain depth, as atoms are bonded further away to other bonds (*e.g.* in sequences of alternating cations and anions) (Figure 4).

3.3. The output of the speciation script consists of two files, *.popul.dat* and *.stat.dat*. The first is the complete list of all the atomic clusters of all the chemical species found in the simulation. Each cluster is written on one line, specifying its chemical formula, the time at which it formed, the time at which it died, its lifetime, a matrix with the list of the atoms forming this cluster.

The second file gives the population analysis with the abundance of each species, both absolute and relative, and the size of the cluster, *i.e.* how many atoms form it. This analysis corresponds to the actual statistics of the coordination polyhedra for the case -r 0; for the case of polymerization, with -r 1 this needs to be treated carefully as some normalization over the relative number of atoms might need to be applied.

4. Compute diffusion coefficients

4.1. Extract the mean square displacements (MSD) of the atoms as a function of time to obtain the self-diffusivity. The standard formula of the MSD is:

$$MSD_\alpha(\tau) = \frac{1}{N_\alpha} \frac{1}{N_{init}} \sum_{1}^{N_\alpha} \sum_{t}^{T/2} [r_\alpha(t+\tau) - r_\alpha(t)]^2$$

where the prefactors are renormalizations. With the MSD tool, there are different ways to analyze the dynamical aspects of the fluids.

NOTE: T is the total time of the simulation and $N_\alpha$ is the number of atoms of type $\alpha$. The initial time $t_0$ is arbitrary and spans the first half of the simulation. $N_{init}$ is the number of initial times. $\tau$ is the width of the time interval over which the MSD are computed; its maximum value is half the time length of the simulation. In typical MSD implementations every window starts at the end of the previous one. But a sparser sampling can speed up the computation of the MSD, without altering the slope of the MSD. For this the i-th window starts at time $t_0(i)$, but the (i+1)-th window starts at time $t_0(i) + \tau + v$, where the value of $v$ is user defined. In a similar way the width of the window is increased in discrete steps defined by the user, as such: $\tau(i) = \tau(i-1) + z$. The values of z ("*horizontal step*") and v ("*vertical step*") are positive or zero; the default for both is 20.

4.2. Compute the MSD using the series of **msd_umd** scripts. Their output is printed in a *.msd.dat* file, where the MSD of each atomic type, atom, or cluster is printed on one column as a function of time.

4.2.1. Compute the average MSD of each atomic type. The MSD are computed for each atom and then averaged for each atomic type. The output file contains one column for each atomic type. At the command line type:

**msd_umd.py -f <UMD_filename> -z <HorizontalJump> -v <VerticalJump> -b <Ballistic>**

4.2.2. Compute the MSD of each individual atom. The MSD are computed for each atom and then averaged for each atomic type. The output file contains one column for each atom in the simulation, and then one column for each atomic type. This feature allows for identifying atoms that diffuse in two different environments, like liquid and gas, or two liquids. At the command line type:

**msd_all_umd.py -f <UMD_filename> -z <HorizontalJump> -v <VerticalJump> -b <Ballistic>**

4.2.3. Compute the MSD of the chemical species. Use the population of clusters identified with the speciation script, and printed in the *.popul.dat* file. The MSD are computed for each individual cluster. The output file contains one column for each cluster. In order to avoid considering large-scale polymers a limit to the size of the cluster can be put; its default is 20 atoms. At the command line type:

**msd_cluster_umd.py -f <UMD_filename> -p <POPUL_filename> -s <Sampling_Frequency> -b <Ballistic> -c <ClusterMaxSize>**

NOTE: Defaults values are: –b 100 –s 1 –c 20

4.3. Plot the MSD using a spreadsheet-based software. In a log-log representation of the MSD vs time identify the slope change. Separate the first part, usually short, which represents the *ballistic* regime, i.e. the conservation of the velocity of atoms after collisions, from the second longer part, which represents the *diffusive* regime, i.e. scattering of the velocity of atoms after collisions.

NOTE: The MSD offer a very strong validation criterion for the quality of the simulations. If the diffusion part of the MSD is not sufficiently long, that is a sign that the simulation is too short, and fails to reach the fluid state in a statistical sense. The minimum requirement for the diffusive part of the MSD highly depends on the system. One can require that all the atoms should change at least once their site in the structure of the melt in order for it to be considered as a fluid[10]. An excellent example with applications in planetary sciences is complex silicate melts at high pressures close to or even below their liquidus line[11]. The Si atoms, the major network-forming cations, switch sites after more than two dozen picoseconds. Simulations shorter than this threshold would be considerably under-sampling the possible configurational space. However as the coordinating anions, namely the O atoms, move faster than the central Si atoms, they can compensate for part of the slow mobility of Si. As such, the entire system could indeed cover a better sampling of the configurational space than assumed only from the Si displacements.

4.4. Plot the MSD using a spreadsheet-based software. Compute the diffusion coefficients as:

$$D = \frac{1}{2\,Z\,t}$$

where Z is the number of degrees of freedom (Z=2 for diffusion in plane, Z=3 for diffusion in space), and t is the time step.

5. **Time correlation functions ensure the passage from microscopic simulation to macroscopic properties for certain properties like transport or spectroscopy.**

5.1. Time correlation functions provide a measure of the inertia of the system. Generally they are expressed as:

$$C(\tau) = \frac{1}{\tau} \sum_{t=0}^{T/2} A(t+\tau)A(t)$$

The A can be a variety of time-dependent variables, such as the atomic positions, atomic velocities, stresses, polarization, etc., each yielding via the Green-Kubo relations[12,13], different physical properties, sometimes after a further transformation.

5.2. Analyze the atomic velocities to obtain the vibrational spectrum of the liquid and an alternative expression of the atomic self-diffusion coefficients.

5.2.1. Run the **vibr_spectrum_umd.py** script. At the command line type:

**vibr_spectrum_umd.py** -f <UMD_filename> -t <temperature>

where –t is the temperature that must be defined by the user. The script prints two files: the *.vels.scf.dat* file with the VAC function for each atomic type, and the *.vibr.dat* file with the vibrational spectrum decomposed on each atomic species and the total value.

5.2.2. Compute the atomic velocity-velocity auto-correlation (VAC) function for each atomic type. Open and read the *vels.scf.dat*.

5.2.3. Fast-Fourier transform the atomic VAC.

5.2.4. Keep the real part of the Fourier VAC. This is what yields the vibrational spectrum, as a function of frequency:

$$I(\omega) = Re[\int_0^\infty m <v(t+\tau)v(t)> e^{-i\omega t}]$$

where $m$ are the atomic masses.

5.2.5. Plot the vibrational spectrum from the *vibr.dat* file using spreadsheet-like software. Identify the finite value at $\omega = 0$ that corresponds to the diffusive character of the fluid and the various peaks of the spectrum at finite frequency. Identify the participation of each atomic type to the vibrational spectrum. The decomposition on atomic types shows that different atoms have different $\omega = 0$ contributions, corresponding to their diffusion coefficients. The general shape of the spectrum is much smoother with fewer features than for a corresponding solid.

5.2.6. Read at the shell the integral over the vibrational spectrum which yields the diffusion coefficients for each atomic species.

NOTE: Thermodynamic properties can be obtained by integration from the vibrational spectrum, but the results should be used with caution because of two approximations: the integration is valid within the quasi-harmonic approximation, which does not necessarily hold at high temperatures; and the gas-like part of the spectrum corresponding to the diffusion needs to be discarded. The integration should then be done *only* over the lattice-like part of the spectrum. But this separation usually requires several further post-processing steps and calculations[14], which are not covered by the present UMD package.

5.3. Run the **viscosity_umd.py** script to analyze the self-correlation of the components stress tensor to estimate the viscosity of the melt.. At the command line type:

**viscosity_umd.py -f <UMD_filename> -i <InitialStep> -s <Sampling_Frequency> -o <frequency_of_origin_shift> -l <max_correlation_timelength>**

NOTE: this feature is exploratory and any results must be taken with caution. First of all thoroughly check the convergence of the viscosity with respect to the length of the simulation.

5.3.1. Deribe the viscosity of the fluid from the self-correlation of the stress tensor[15] as:

$$\eta = \frac{V}{3k_BT} \sum_{i,j} \int_0^\infty <\sigma_{ij}(t+\tau)\sigma_{ij}(t)> d\tau$$

where V and T are the volume and the temperature respectively, $k_B$ is the Boltzmann constant and $\sigma_{ij}$ the *ij* off-diagonal component of the stress-tensor, expressed in Cartesian coordinates.

5.3.2. However, the stress-tensor auto-correlation function can be rather noisy because of the finite size and the finite duration of the simulations. We can however try to use a more adequate fit to obtain a more robust estimate of the viscosity[15,16]. For the auto-correlation function of the stress tensor, the following functional form [15,16] yields good results:

$$\frac{1}{3}\sum_{i,j} <\sigma_{ij}(t+\tau)\sigma_{ij}(t)> = Ae^{-\frac{t}{\tau_1}} + Be^{-\frac{t}{\tau_2}}\sin(\omega t)$$

where A, B, $\tau_1$, $\tau_2$, and $\omega$ are fit parameters. After integrating, the expression for the viscosity becomes:

$$\frac{\eta k_B T}{V} = A\tau_1 + B\frac{\tau_2}{(1+\omega^2\tau_2^2)}$$

6. Finally, two scripts can quickly provide a thorough analysis of the major thermodynamic parameters stemming from the simulations.

6.1. Run the **averages.py** to extract from the *umd* files the average values and the spread (as standard deviation) for pressure, temperature, density, and internal energy. At the command line type:

**averages.py -f <UMD_filename> -s <Sampling_Frequency>**

with –s 0 as default.

6.2. An alternative formulation of the error is the statistical error of the average, computed using the blocking method. There are various flavors of this method. Following the work of Allen and Tildesley[2], it is common to average over sequences of time blocks, of increasingly longer length, and estimate the standard deviation with respect to the arithmetic average[17]. Convergence may be reached in the limit of many- and long-*enough* block sizes, when the sampling is uncorrelated. Though the actual threshold value for the convergence usually needs to be chosen manually.

An equivalent formulation that we adopt here is to use the halving method[18]. Starting with the initial data sample, at each step $k$ the number the samples is halved by averaging over every two corresponding consecutive samples from the previous step $k-1$:

$$S_i^k = \frac{S_{2i}^{k-1} + S_{2i+1}^{k-1}}{2}$$

6.2.1. Run the **fullaverages.py** script to extract the complete statistical analysis, including the error on the mean. At the command line type:

**fullaverages.py -s <Sampling_Frequency> -u <units>**

NOTE: The script is automatized to the point of searching for all the *.umd.dat* files in the current directory and performing the analysis for all of them. Defaults are –s 0 –u 0. For -u 0 the output is minimal, and for -u 1 output is in full, with several alternative units printed out. This script requires a graphical support, as it creates a graphical image for checking the convergence for estimating the error on the mean.

**REPRESENTATIVE RESULTS:**

Pyrolite is a model multi-component silicate melt (0.5Na$_2$O 2CaO 1.5Al$_2$O$_3$ 4FeO 30MgO 24SiO$_2$) that best approximates the composition of the bulk silicate Earth – the geochemical average or our entire planet, except for its iron-based core[19]. The early Earth was dominated by a series of large-scale melting events[20], the last one that might have engulfed the entire planet, after its condensation for the protolunar disk[21]. Pyrolite represents the best approximant to the composition of such planetary-scale magma oceans. Consequently, we extensively studied the physical properties of pyrolite melt in the 3000-5000K temperature range and 0-150 GPa pressure range from ab initio molecular-dynamics simulations in the VASP implementation. These thermodynamic conditions characterize entirely the most extreme Earth's magma ocean conditions. Our study is an excellent example of a successful use of the UMD package for the entire in-depth analysis of the melts[22]. We computed the distribution and the average bond lengths, we traced the changes in cation-oxygen coordination, and compared our results to previous experimental and computational studies on amorphous silicates of various compositions. Our in-depth analysis helped decompose standard coordination numbers into their basic constituents, outline the presence of exotic coordination polyhedra in the melt, and extract lifetimes for all the coordination polyhedra. It also outlined the importance of sampling in simulations in terms of both length of the trajectory but also number of atoms present in the system that is modeled. As post-processing, the UMD analysis is independent of these factors, however they should be taken into account when interpreting the results provided by the UMD package. We show here a few examples of how the UMD package can be used to extract several characteristic features of the melts, with an application to molten pyrolite.

The Si-O pair distribution function obtained from the gofrs_umd.py script shows that the radius of the first coordination sphere, which is the first minimum of the g(r) function, lies around 2.5 angstroms at T=3000K and P=4.6 GPa. The maximum of the g(r) lies at 1.635 angstroms – this is the best approximation to the bend length. The long tail is due to the temperature. Using this limit as the Si-O bond distance, the speciation analysis shows that $SiO_4$ units, which can last for up to a few picoseconds, dominate the melt (Figure 5). There is an important part of the melt that shows partial polymerization, as reflected by the presence of dimers like $Si_2O_7$, and trimers like $Si_3O_x$ units. Their corresponding lifetime is on the order of the picosecond. Higher-order polymers all live considerably less.

The different values of the vertical and horizontal steps, defined by the –z and –v flags above, yield various samplings of the mean-square displacements (Figure 5). Even large values of z and v are enough to define the slopes and thus the diffusion coefficients of the different atoms. The gain in time for the post-processing is remarkable when going to large values of z and v.

Finally the atomic VAC functions yield the vibrational spectrum of the melt. Figure 6 shows the spectrum at the same pressure and temperature conditions as above. We represent the contributions of Mg, Si, and O atoms, as well as the total value. At zero frequency there is a finite value of the spectrum, which corresponds to the diffusion character of the melt. Extraction of the thermodynamic properties from the vibrational spectrum need to remove this gas-like diffusive character from zero but also to properly take into account its decay at higher frequencies.

**FIGURE AND TABLE LEGENDS:**

Figure 1. Flowchart of the UMD library. Physical properties are in blue, major Python scripts and their options in red.

Figure 2. The beginning of the umd file describing the simulation of liquid pyrolite at 4.6 GPa and 3000K. The header is followed by the description of each snapshot. Each property is written on one line, containing the name of the physical property, the value(s), and the units, all separated by spaces.

Figure 3. For each atom of one species (for example red) all the atoms of the coordinating species (for example grey and/or red) are counted as a function of distance (a). The resulting distance distribution graph (b) for each snapshot, which at this stage is only a collection of delta functions, is then averaged over all the atoms and all the snapshots and weighted by the ideal gas distribution to generate the pair distribution function (c) that is continuous. The first minimum of the g(r) is the radius of the first coordination sphere, used later on in the speciation analysis.

Figure 4. Typical analysis and speciation defined as polymerization. The coordination polyhedra are defined using interatomic distances. All atoms at a distance smaller than a specified radius are considered to be bonded. Here the threshold corresponds to the first coordination sphere

(the light red circles), defined in Figure 1. Polymerization and thus chemical species are obtained from the networks of the bonded atoms. Note the central $Red_1Grey_2$ cluster, which is isolated from the other atoms, which form an infinite polymer.

Figure 5. Example of using the UMD package on a multi-component silicate melt: lifetime of the Si-O chemical species identified in the melt at 4.6 GPa and 3000K. The labels mark the $SiO_3$, $SiO_4$, and $SiO_5$ monomers and the various $Si_xO_y$ polymers.

Figure 6. Example of using the UMD package on a multi-component silicate melt: the sampling of the mean-square displacements with various horizontal and vertical steps, z and v, yield consistent results. Solid circles: -z 50 –v 50. Open circles: -z 250 –v 500.

Figure 7. Example of using the UMD package on a multi-component silicate melt: the real part of the Fourier transform of the atomic velocity-velocity self-correlation function yields the vibrational spectrum. Fluids have a non-zero gas-like diffusive character at zero frequency.

Table 1. Most common flags used in the UMD package and their most common significance.

**DISCUSSION:**

By its conception the UMD package is designed to work better with ab initio simulations, where the number of snapshots is typically limited to tens to hundreds of thousands of snapshots, with a few hundred atoms per unit cell. Larger simulations are also tractable provided the machine on which the post-processing runs has enough active memory resources. The code distinguishes itself by the variety of properties it can compute and by its open-source license.

The umd.dat files are appropriate to the ensembles that preserve the number of particles unchanged throughout the simulation. The UMD package can read files stemming from calculations where the shape and volume of the simulation box varies. These cover all the most common calculations, like NVT and NPT, where the number of particles, N, temperature T, volume, V, and/or pressure, P, are kept constant.

For the time begin the pair distribution function as well as all the scripts needing to estimate the interatomic distances, like the speciation scripts, work only for orthogonal unit cells, meaning for cubic, tetragonal, and orthorhombic cells, where the angles between the axis are 90 degrees.

The major lines of development for version 2.0 are removal of the orthogonality restriction for distances and adding more features for the speciation scripts: to analyze individual chemical bonds, to analyze the interatomic angles, and to implement the second coordination sphere. With help from a separate external collaboration we are working on porting the code on GPU for faster analysis on larger systems.

**ACKNOWLEDGMENTS:**

This work was supported by the European Research Council (ERC) under the European Union Horizon 2020 research and innovation program (grant agreement number 681818 IMPACT), by the Extreme Physics and Chemistry Directorate of the Deep Carbon Observatory, and by the Research Council of Norway through its Centres of Excellence funding scheme, project number 223272. We acknowledge access to the GENCI supercomputers through the stl2816 series of eDARI computing grants, to the Irene AMD supercomputer through the PRACE RA4947 project, and the Fram supercomputer through the UNINETT Sigma2 NN9697K. FS was supported by a Marie Skłodowska-Curie project (grant agreement ABISSE No.750901).

**DISCLOSURES:**

The authors have nothing to disclose.

**FIGURE AND TABLE LEGENDS:**

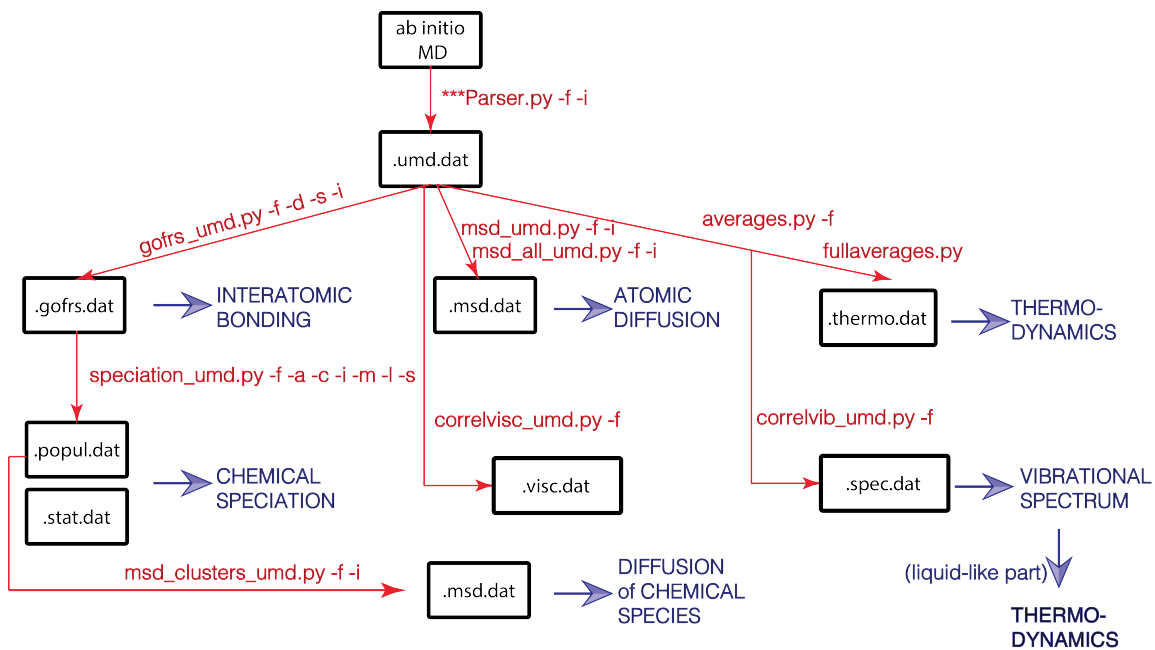

Figure 1. Flowchart of the UMD library. Physical properties are in blue, major python scripts and their options in red.

```
natom 153
ntypat 7
types 1 2 4 30 3 24 89
elements Na Ca Fe Mg Al Si O
typat 0 1 1 2 2 2 2 3 3 3 3 3 3 3 3 3 3 3 3 3 3 3 3 3 3 3 3 3 3 3 3 3 3 3 3 4 4 5 5 5 5 5 5 5 5 5 5 5 5 5 5 5 5 5 5 5 5 5 5 5 5 5 5 6 6 6 6 6 6 6 6 6 6 6 6 6
6 6 6 6 6 6 6 6 6 6 6 6 6 6 6 6 6 6 6 6 6 6 6 6 6 6 6 6 6 6 6 6 6 6 6 6 6 6 6 6 6 6 6 6 6 6 6 6 6 6 6 6 6 6 6 6 6 6 6 6 6 6 6 6 6 6 6 6 6 6 6 6 6 6 6 6 6 6 6

timestep 1.0 fs
time 0.0 fs
InternalEnergy -983.52058 eV
Temperature 3010.0 K
Pressure 9.426 GPa
StressTensor 96.4 72.78 113.61 14.03 -4.27 -9.76  GPa
acell 12.0 12.0 12.0 Angstroms
rprim_a 1.0  0.0  0.0 Angstroms
rprim_b 0.0  1.0  0.0 Angstroms
rprim_c 0.0  0.0  1.0 Angstroms
rprimd_a 12.0  0.0  0.0 Angstroms
rprimd_b 0.0  12.0  0.0 Angstroms
rprimd_c 0.0  0.0  12.0 Angstroms
atoms: reduced*3 cartesian*3(A) abs.diff.*3(A) velocity*3(A/fs) force*3(eV/A)
0.30463 0.52016 0.14339 3.65551 6.24194 1.72064 3.65551 6.24194 1.72064 -0.00222 0.0092 0.00297 -0.24013 -1.42767 -0.88373
0.48556 0.65381 0.90411 5.82672 7.84568 10.84929 5.82672 7.84568 10.84929 -0.00576 0.0076 -0.00125 2.45718 0.71688 1.46513
0.18827 0.30872 0.1126 2.25923 3.70465 1.35122 2.25923 3.70465 1.35122 0.01147 0.00074 -0.00436 -0.93183 1.59251 -1.40121
0.77668 0.7787 0.69676 9.32019 9.34434 8.36111 9.32019 9.34434 8.36111 0.0014 0.00765 0.00471 0.40742 -1.70923 1.47656
0.61119 0.93599 0.78855 7.33426 11.23183 9.46266 7.33426 11.23183 9.46266 0.00079 0.00487 0.00063 1.65204 3.2221 0.74208
0.12942 0.67933 0.70957 1.55302 8.15197 8.51484 1.55302 8.15197 8.51484 0.00028 0.00766 -7e-05 0.14531 -0.11801 2.44219
0.54902 0.91138 0.52596 6.58829 10.93659 6.31148 6.58829 10.93659 6.31148 -0.00217 0.00417 -0.00105 -6.54642 -3.51328 3.05384
0.7713 0.28643 0.55361 9.25556 3.43715 6.64332 9.25556 3.43715 6.64332 0.03582 -0.00389 0.00396 0.62891 -1.55255 -0.40697
0.19748 0.65517 0.9768 2.36973 7.86208 11.72166 2.36973 7.86208 11.72166 0.01141 0.00721 0.00429 -0.83738 -2.4069 1.34372
```

Figure 2. The beginning of the umd file describing the simulation of liquid pyrolite at 4.6 GPa and 3000K. The header is followed by the description of each snapshot. Each property is written on one line, containing the name of the physical property, the value(s), and the units, all separated by spaces.

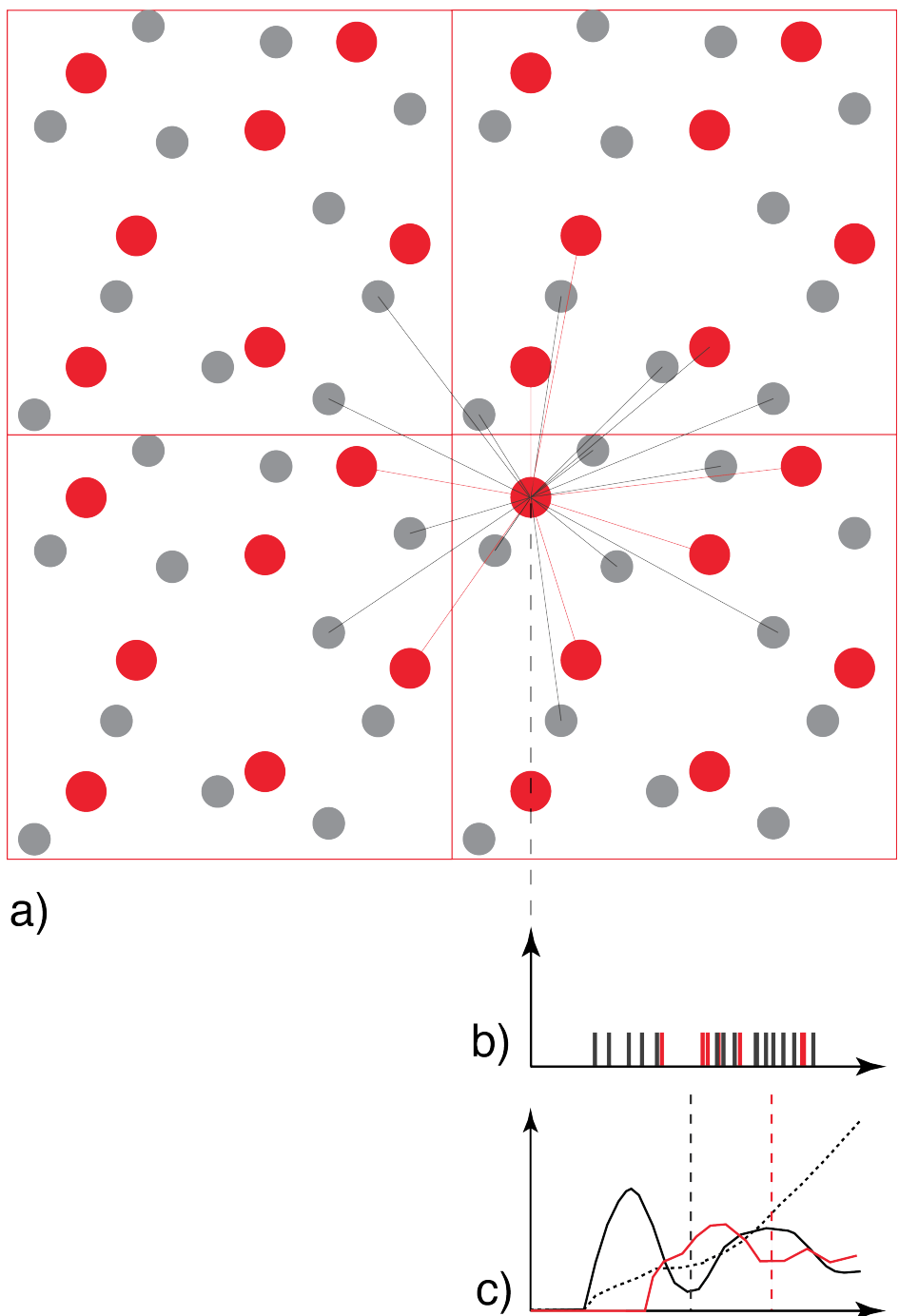

Figure 3. For each atom of one species (for example red) all the atoms of the coordinating species (for example grey and/or red) are counted as a function of distance (a). The resulting distance distribution graph (b) for each snapshot, which at this stage is only a collection of delta functions, is then averaged over all the atoms and all the snapshots and weighted by the ideal gas distribution to generate the pair distribution function (c) that is continuous. The first minimum of the g(r) is the radius of the first coordination sphere, used later on in the speciation analysis.

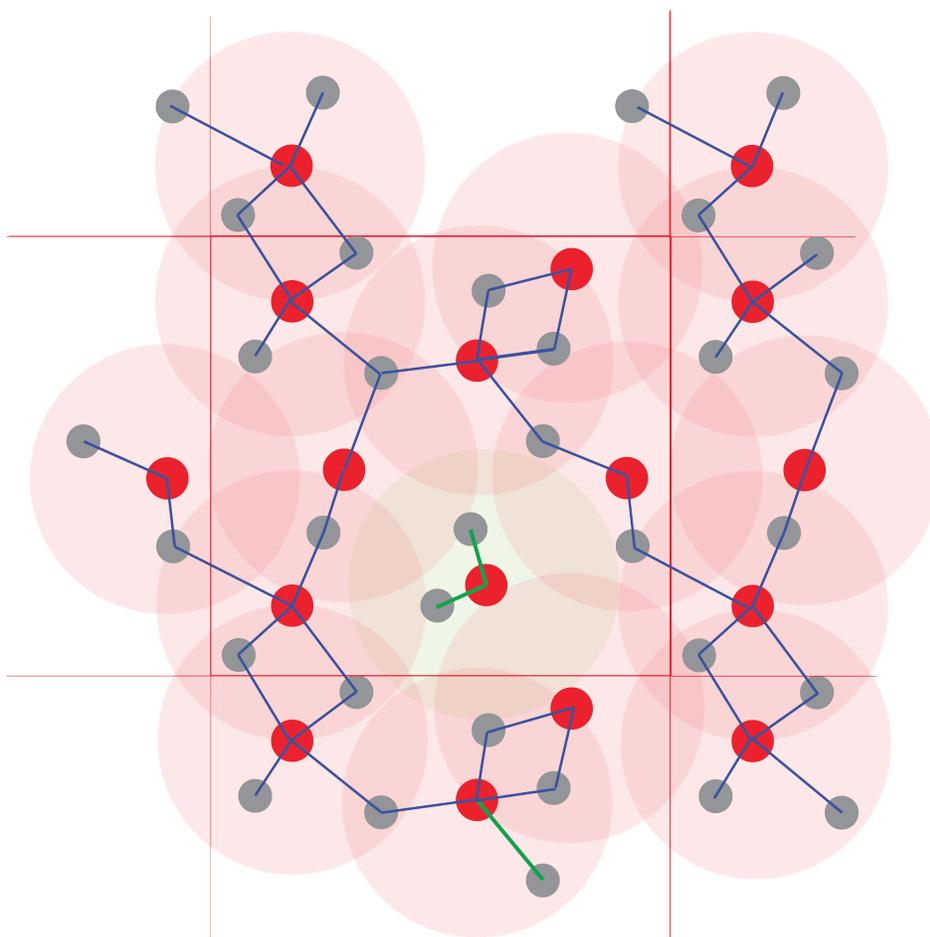

Figure 4. Typical analysis and speciation defined as polymerization. The coordination polyhedra are defined using interatomic distances. All atoms at a distance smaller than a specified radius are considered to be bonded. Here the threshold corresponds to the first coordination sphere (the light red circles), defined in Figure 1. Polymerization and thus chemical species are obtained from the networks of the bonded atoms. Note the central $Red_1Grey_2$ cluster, which is isolated from the other atoms, which form an infinite polymer.

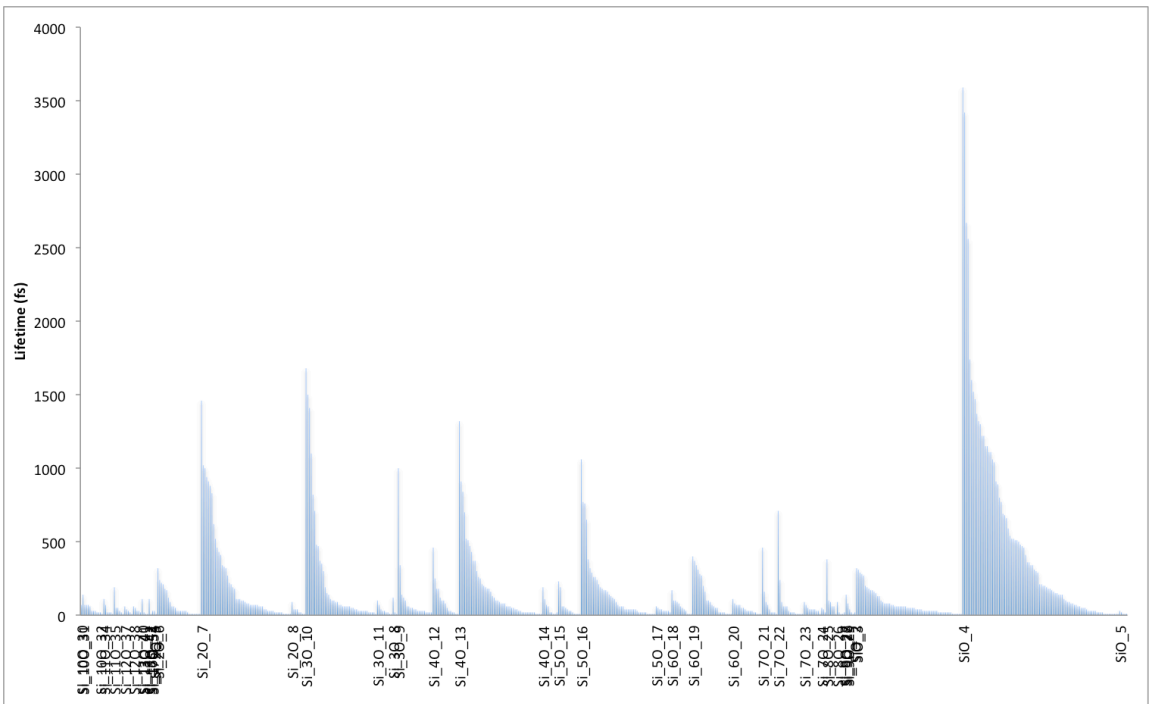

Figure 5. Example of using the UMD package on a multi-component silicate melt: lifetime of all $Si_xO_y$ chemical species identified in the melt at 4.6 GPa and 3000K.

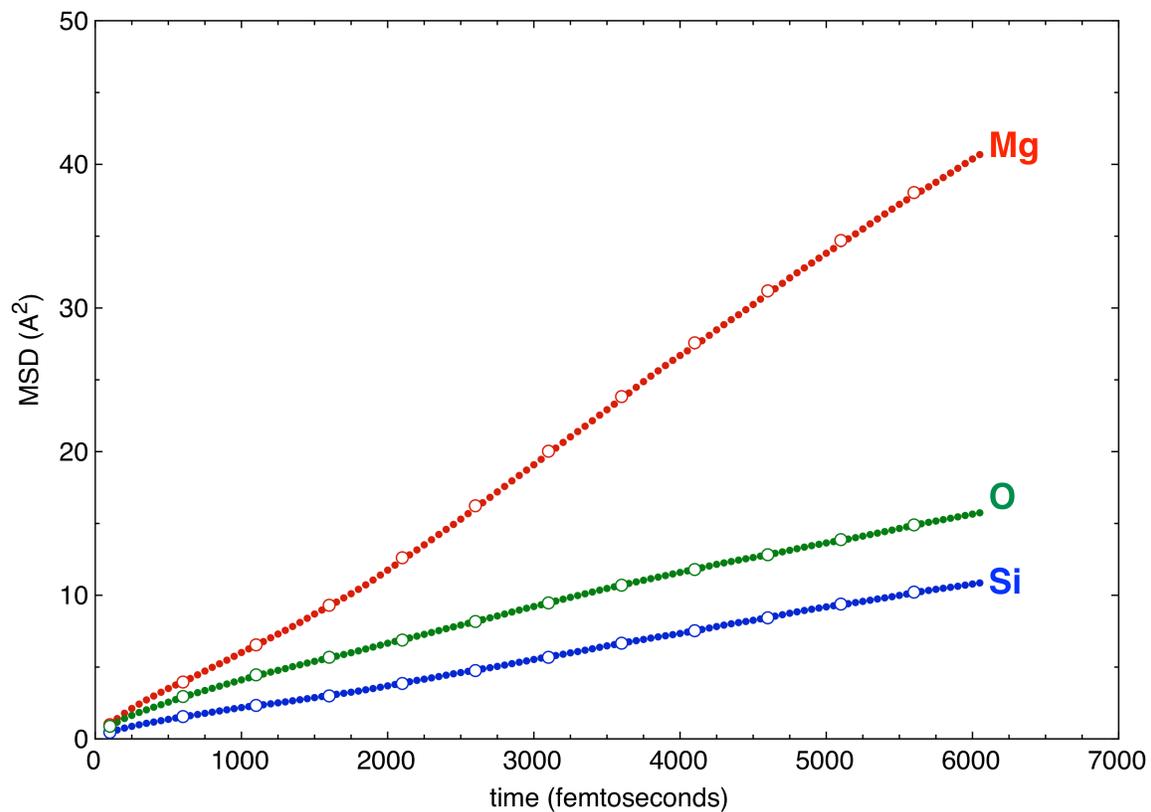

Figure 6. Example of using the UMD package on a multi-component silicate melt: the sampling of the mean-square displacements with various horizontal and vertical steps, z and v, yield consistent results. Solid circles: -z 50 –v 50. Open circles: -z 250 –v 500.

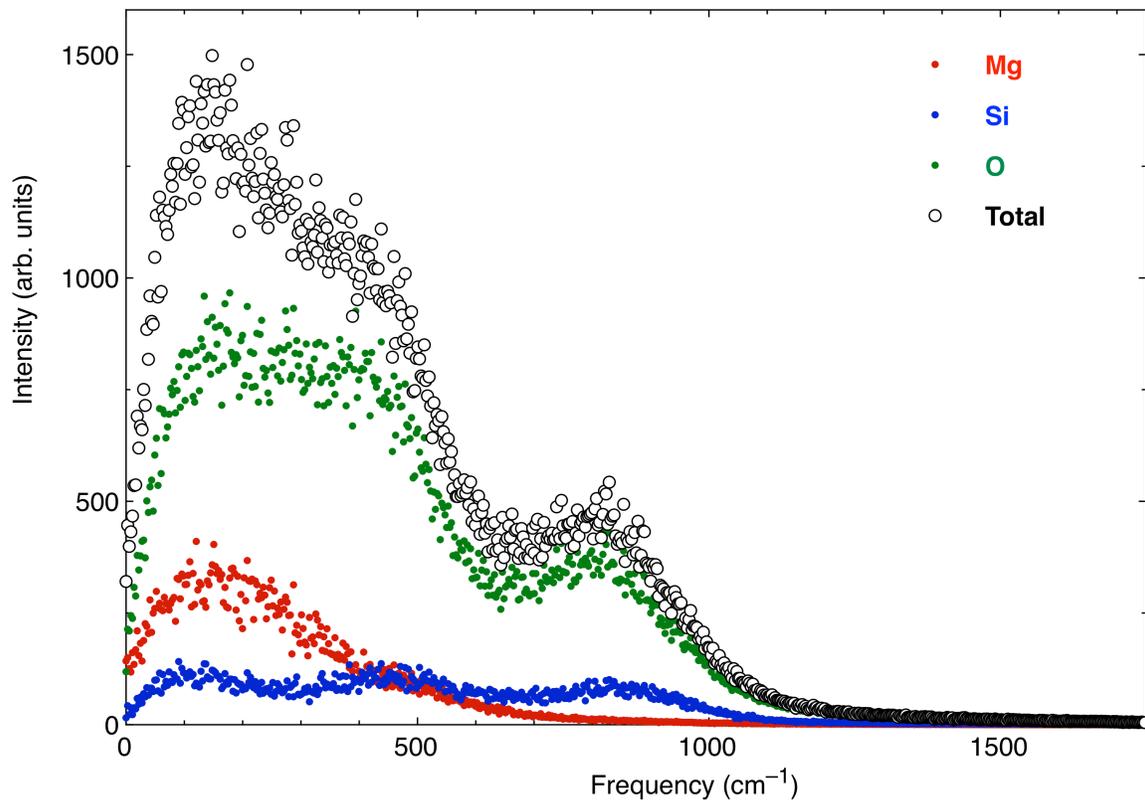

Figure 7. Example of using the UMD package on a multi-component silicate melt: the real part of the Fourier transform of the atomic velocity-velocity self-correlation function yields the vibrational spectrum. Fluids have a non-zero gas-like diffusive character at zero frequency.